
%
\input phyzzx
\tolerance=1000
\sequentialequations
\def\rl{\rightline}

\def\r#1{$\bf#1$}

\def\t1{{\tilde 1}}

\def\AEF{A.E. Faraggi}

\def\NPB#1#2#3{Nucl. Phys. B {\bf#1} (19#2) #3}
\def\PLB#1#2#3{Phys. Lett. B {\bf#1} (19#2) #3}
\def\PRD#1#2#3{Phys. Rev. D {\bf#1} (19#2) #3}

\def\IJMP#1#2#3{Int. J. Mod. Phys. A {\bf#1} (19#2) #3}

\def\l{\langle}
\def\r{\rangle}

\REF\RP{S. Dimopoulos and H. Georgi, \NPB{193}{81}{150}.}
\REF\BAR{S. Weinberg, \PRD{26}{82}{287}; N. Sakai and Y. Yanagida, \NPB{197}
{82}{533}.}
\REF\LEP{L. J. Hall and M. Suzuki, \NPB{231}{84}{419}.}
\REF\MOD{\AEF, \PLB{278}{92}{131}.}
\REF\SLM{\AEF, \PLB{274}{92}{47}; \NPB{387}{92}{289}.}
\REF\DSW{M. Dine, N. Seiberg and E. Witten, \NPB{289}{87}{585};
J.J. Atick, L.J. Dixon and A. Sen, \NPB{292}{87}{109};
S. Cecotti, S. Ferrara and M. Villasante, \IJMP{2}{87}{1839}.}
\REF\FFF{I. Antoniadis, C. Bachas, and C. Kounnas, \NPB{289}{87}{87};
I. Antoniadis and C. Bachas, \NPB{298}{88}{586};
5H. Kawai, D.C. Lewellen, and S.H.-H. Tye,
Phys. Rev. Lett. {\bf57} (1986) 1832;
Phys. Rev. D {\bf 34} (1986) 3794;
Nucl. Phys. B {\bf 288} (1987) 1.}
\REF\KLN{S. Kalara, J. Lopez and D.V. Nanopoulos, \PLB{245}{91}{421};
\NPB{353}{91}{650}.}
\REF\BAN{T. Banks and L. Dixon, \NPB{307}{88}{93}.}
\REF\NRT{\AEF, \NPB{403}{93}{101}.}
\REF\CKM{A. E. Faraggi and E. Halyo, \PLB{307}{93}{305}; WIS--93/34/MAR--PH.
to appear in Nucl. Phys. B.}
\REF\LQ{E. Halyo, WIS--93/114/DEC--PH.}
\REF\FM{\AEF, \NPB{407}{93}{57}.}
\REF\PRO{E. Halyo, WIS preprint in preparation.}

\singlespace
\rl{WIS--94/4/JAN--PH}
\rl{\today}
\rl{T}
\pagenumber=0
\normalspace
\smallskip
\titlestyle{\bf{R Parity in Standard--like Superstring Models}}
\smallskip
\author{Edi Halyo{\footnote*{e--mail address: jphalyo@weizmann.bitnet}}}
\smallskip
\centerline {Department of Particle Physics}
\centerline {Weizmann Institute of Science}
\centerline {Rehovot 76100, Israel}
\vskip 6 cm
\titlestyle{\bf ABSTRACT}

We investigate the R symmetries of standard--like superstring models. At the
level of the cubic superpotential there are three global $U(1)$ R symmetries.
These are broken explicitly by $N>3$ terms in the superpotential
and spontaneously by scalar VEVs necessary to preserve supersymmetry at $M_P$.
A $Z_2$ discrete symmetry remains but is equivalent to fermion number modulo
2. These models possess an
effective R parity which arises from the interplay between the gauged
$U(1)_{B-L}$ and $U(1)_{r_{j+3}}$.

\singlespace
\vskip 0.5cm
\endpage
\normalspace

\centerline{\bf 1. Introduction}

Supersymmetric extensions of the Standard Model (SM), whatever their
origin, may contain dangerous terms which violate baryon
number ($B$) and lepton number ($L$) at unacceptably large rates. This is
because, contrary to the SM case, in these models the SM gauge symmetry does
not result in accidental (global) $B$ and $L$ symmetries. For example,
the superpotential of the minimally supersymmetric Standard Model (MSSM)
may contain (dimension four) terms like
[\RP,\BAR]
$$c_1 Q_i L_j d_k + c_2 u_i d_j d_k +c_3 L_i L_j e_k ,\eqno(1)$$
where $i,j,k$ are generation indices. The first two operators induce an
unacceptably large proton decay rate unless $c_1 c_2<10^{-26}$ [\BAR]
(for squark and slepton masses of at most a TeV). The third operator induces
$L$ violating rare events such as $\mu \to e^+e^-e^+$ unless $c_3$ is
sufficiently suppressed [\LEP]. It is difficult to explain such small
couplings without resorting to a symmetry that protects them. An elegant way of
obtaining such small (or vanishing) couplings is to impose an R parity on
the model [\RP,\BAR]. This is a $Z_2$ symmetry which does not
commute with supersymmetry (SUSY) and under which the superfields have the
following charges: $Q_i,L_i,u_i,d_i,e_i$ (and $N_i$, the right--handed neutrino
if it exists) all have $-1$ (odd), $H_1,H_2$ and the vector superfields
have $+1$ (even). In other words, under R parity all SM states are even
whereas all their superpartners are odd. R parity, as defined above, eliminates
the dangerous
terms in Eq. (1) without affecting the usual terms in the MSSM Lagrangian.

The above considerations hold not only for MSSM but for any supersymmetric
extension of the SM. In particular, any superstring model which reduces to MSSM
or some extension of it at the TeV scale must somehow suppress the terms in
Eq. (1) enough so that constraints from $B$ and $L$ violation are satisfied.
In this letter, we examine the R parity in standard--like superstring
models[\MOD,\SLM]. We find that there are three continous $U(1)$ R symmetries
at the cubic level of the superpotential. These are broken explicitly by higher
order ($N>3$) terms in the superpotential and spontaneously by the scalar VEVs
which are necessary in order to preserve SUSY around the Planck scale. Then,
only a discrete $Z_2$ subgroup of the three $U(1)$'s survive. We find that
this $Z_2$ is not the R parity usually assumed in MSSM and does not eliminate
any
of the terms in Eq. (1). In fact, it is equivalent to fermion number modulo 2.
On the other hand, standard--like superstring models possess an effective
R parity which arises from the particular charges of observable and hidden
sector states under gauged $U(1)_{B-L}$ and $U(1)_{r_{j+3}}$.

The standard--like superstring models that we consider have the following
properties [\MOD,\SLM]:

1. $N=1$ space--time SUSY.

2. A $SU(3)_C\times SU(2)_L\times {U(1)^n}\times$hidden gauge group.

3. Three generations of chiral fermions
and their superpartners, with the correct quantum numbers
under ${SU(3)_C\times SU(2)_L\times U(1)_Y}$.

4. Higgs doublets that can  produce realistic electro--weak symmetry breaking.

5. Anomaly cancellation, apart from a single ``anomalous" U(1)
which is  canceled by  application of the
Dine--Seiberg--Witten (DSW) mechanism [\DSW].

The standard--like superstring models are constructed in the four
dimensional free fermionic formulation [\FFF].
The models are generated by a basis of eight boundary condition vectors
for all world--sheet fermions [\MOD,\SLM].
The observable and hidden gauge groups after application
of the generalized GSO projections are $SU(3)_C\times U(1)_C\times
 SU(2)_L\times U(1)_L\times U(1)^6${\footnote*{$U(1)_C={3\over 2}U(1)_{B-L}$
and $U(1)_L=2U(1)_{T_{3_R}}$.}}
and $SU(5)_H\times SU(3)_H\times U(1)^2$, respectively.
The weak hypercharge is given by
$U(1)_Y={1\over 3}U(1)_C + {1\over 2}U(1)_L$ and has the standard $SO(10)$
embedding. The orthogonal
combination is given by $U(1)_{Z^\prime}= U(1)_C - U(1)_L$.

The models have six right--handed gauge $U(1)_r$ symmetries which correspond to
the right--handed world--sheet currents $\bar \eta^j \bar \eta^{j^*}$ (j=1,2,3)
and $\bar y^3 \bar y^6, \bar y^1 \bar \omega^5, \bar \omega^2 \bar \omega^4$.
At the level of the cubic superpotential, there is a left--handed global
$U(1)_{\ell}$
symmetry for every right--handed gauge $U(1)_r$. The six left--handed global
symmetries correspond to the left--handed world--sheet currents $\chi^{12},
\chi^{34}, \chi^{56}$ and $ y^3 y^6, y^1 \omega^5, \omega^2 \omega^4$.
We concentrate on the first three of these since their sum gives the $U(1)$ of
the $N=2$ world-sheet SUSY [\KLN]
algebra which means that $\chi^{12},\chi^{34},\chi^{56}$ are SUSY charges.
This can also be seen from the basis vector S
$$S=({\underbrace{1,\cdots,1}_{{\psi^\mu},{\chi^{12,34,56}}}},0,\cdots,0
\vert 0,\cdots,0). \eqno(2)$$
S plays the part of the SUSY generator in the sense that the SUSY
partners of the states from any sector $\alpha$ are given by the sector
$S+\alpha$. The three $\chi^{ij}$'s make up the SUSY generator S. Therefore
different (scalar or fermionic) components of superfields will have different
$Q_{\ell_i}$ ($i=1,2,3$) charges exactly as for R symmetries.
We will see that the world--sheet currents $\chi^{12},\chi^{34},\chi^{56}$
correspond to space--time global $U(1)$ R symmetries at the
cubic level of the superpotential. Of course, due to a well--known theorem
[\BAN], there are no continous global symmetries
in strings. These global $U(1)$ R symmetries are broken
explicitly by $N>3$ terms in the superpotential and spontaneously by scalar
VEVs required by the stability of the SUSY vacuum near the Planck mass. As a
result there remains only a $Z_2$ subgroup (R parity of the model) which is
equivalent to fermion number modulo 2.

A generic standard--like superstring model including
the complete massless spectrum with the gauge quantum numbers and the cubic
superpotential were presented in Ref. [\MOD] and will not be repeated here.
We use the notation of Ref. [\MOD] throughout this letter.
The massless states with their R charges given by $U(1)_{\ell_i}$ are:

(a) The sectors $b_{1,2,3}$ which give the three chiral generations.
States from each sector $b_i$ have $Q_{\ell_i}=1/2$. The superpartners of the
chiral fermions in sector $b_i$ come from the sectors $S+b_i$ with
$Q_{\ell_j}=Q_{\ell_k}=-1/2$ where $i \not=j \not=k$.

(b) The sector $b_1+b_2+\alpha+\beta$ gives the following scalars: a weak
doublet $h_{45}$, a color triplet $D_{45}$ and $SO(10)$ singlets $\Phi_{45},
\Phi_1^\pm, \Phi_2^\pm, \Phi_3^\pm$ all with the R charges $Q_{\ell_1}=
Q_{\ell_2}=-1/2$. The superpartners of these states have $Q_{\ell_3}=1/2$.

(c) The Neveu--Shwarz sector gives the graviton, the dilaton, the antisymmetric
tensor and the gauge bosons of the model all with vanishing R charges. In
addition this sector gives the weak doublets $h_{1,2,3}$ and the singlets
$\Phi_{23},\Phi_{13},\Phi_{12}$ and $\xi_1,\xi_2,\xi_3$ with the R charges:
$$\eqalignno{&Q_{\ell_1}(h_1)=Q_{\ell_1}(\Phi_{23})=Q_{\ell_1}(\xi_1)=-1, &(3a)
\cr
&Q_{\ell_2}(h_2)=Q_{\ell_2}(\Phi_{13})=Q_{\ell_2}(\xi_2)=-1, &(3b) \cr
&Q_{\ell_3}(h_3)=Q_{\ell_3}(\Phi_{12})=Q_{\ell_3}(\xi_3)=-1, &(3c)}$$
and all other charges vanish. The supersymmetric partners of these are obtained
by substituting the degenerate vacuum of the S sector instead of the
Neveu--Shwarz vacuum. The superpartners of the graviton, the dilaton, the
antisymmetric tensor and the gauge bosons have
$Q_{\ell_1}=Q_{\ell_2}=Q_{\ell_3}=1/2$. The superpartners of other
Neveu--Shwarz states with $Q_{\ell_i}=-1$ have $Q_{\ell_j}=Q_{\ell_k}=1/2,
\quad Q_{\ell_i}=-1/2$ where again $i \not =j \not =k$.

The barred conterparts of the above states (in the notation of Ref. [\MOD])
have the same $U(1)_{\ell_i}$ charges as the unbarred states.
In addition there are hidden sector states from sectors $b_i+2 \gamma$ and
$b_{1,2}+b_3+\alpha \pm \gamma+(I)$ which have nonzero $Q_{\ell_i}$
[\MOD,\SLM]. None of
the results we obtain change if these states are taken into account and
therefore we will neglect them in the following.

We see that the states of the model and their superpartners have different
charges under the three R symmetries. This is expected since the particles and
their SUSY counterparts appear in the same superfields and the superspace
parameter $\theta$ carries R charges. We find the R charges of $\theta$
to be $Q_{\ell_1}=Q_{\ell_2}=Q_{\ell_3}=1/2$ by inspecting the
difference between the R charges of the different components of superfields.
The cubic superpotential is obtained by calculating the cubic world--sheet
correlators $A_3 \sim \l V_1^f V_2^f V_3^b \r$ using the rules of Ref. [\KLN]
and is given in Ref. [\MOD]. (Here $V_i^f$ $(V_i^b)$ are the fermionic (scalar)
components of the vertex operators.) All of the terms in the cubic
superpotential have the R
charges $Q_{\ell_1}=Q_{\ell_2}=Q_{\ell_3}=-1$ so that the integral over
superspace, $\int d^2 \theta W$, is R invariant. On the other hand,
string selection rules impose the space--time (or field theory) selection rules
[\KLN]
$$\sum Q_{\ell_1}= \sum Q_{\ell_2}= \sum Q_{\ell_3}=0, \eqno(4)$$
on the F terms obtained at the cubic level.
These are seen, at the field theory level, as symmetries of the cubic
superpotential. Since there are no gauge bosons corresponding to these
symmetries in the massless spectrum, $Q_{\ell_i}$ are global symmetries. (The
corresponding gauge bosons are necessarily projected out by the generalized GSO
projection in order to get $N=1$ space--time SUSY.)

We now show that the three $U(1)$ R symmetries are explicitly and spontaneously
broken. The R symmetries are broken explicitly by $N>3$
nonrenormalizable contributions to the superpotential. These terms are
obtained by calculating correlators between vertex operators [\KLN]
$$A_N\sim\langle V_1^fV_2^fV_3^b\cdot\cdot\cdot V_N^b\rangle. \eqno(5)$$
The nonvanishing terms are obtained by
applying the rules of Ref. [\KLN]. In order to obtain the correct ghost charge,
for an order $N$ term, $N-3$ vertex operators are picture changed by taking
$$V_{q+1}(z)=\lim_{w \to z}e^c(w)T_F(w)V_q(z), \eqno(6)$$
where $T_F$ is the world--sheet super current given by
$$T_F=\psi^{\mu}\partial_{\mu}X+i\sum_{I=1}^6 \chi^I y^I \omega^I=T_F^0+
T_F^{-1}+T_F^{+1}, \eqno(7)$$
with
$$T_F^{-1}=e^{-i\chi^{12}}\tau_{_{12}}+e^{-i\chi^{34}}\tau_{_{34}}+
e^{-i\chi^{56}}\tau_{_{56}}  \qquad  T_F^{-1}=(T_F^{+1})^*, \eqno(8)$$
where $\tau_{ij}={i\over \sqrt 2}(y^i\omega^i+iy^j\omega^j)$ and
$e^{i\chi^{ij}}={1\over \sqrt2}(\chi^i+i\chi^j)$.
It can be shown that only the $T_F^{+1}$ piece of $T_F$ contributes to $A_N$
[\KLN] and therefore each picture changing adds one unit to either one of
one of the R charges. Now, the string selection rules require that
$\sum Q_{\ell_i}=0$ for $i=1,2,3$
are satisfied after all picture changings have been performed. As a result, a
generic order $N$ term which requires $N-3$
picture changings does not satisfy the conservation rules given in Eq. (4) with
the R charges given before. Conversely, only order $N$ terms with $\sum
Q_{\ell_1}+Q_{\ell_2}+Q_{\ell_3}=3-N$ survive the string selection rules.
Thus, $N>3$ terms, in general, break the R symmetries explicitly.
For example at order $N=5$ we have the quark mass terms [\NRT,\CKM]
$$\eqalignno{&u_1 Q_1 \bar h_1 \bar \Phi_i^+ \bar \Phi_i^-, &(9a) \cr
             &d_1 Q_1 h_{45} \Phi_1^+ \xi_2, &(9b) \cr
             &d_2 Q_2 h_{45} \bar \Phi_2^- \xi_1, &(9c)}$$
which break $Q_{\ell_1}$ and $Q_{\ell_2}$ explicitly.
$Q_{\ell_3}$ is explicitly broken by the leptoquark--quark mixing term [\LQ]
$$d_3 D_{45} N_3 \Phi_{13} \Phi_3^+ (\xi_1+\xi_2). \eqno(10)$$

As stated above, the massless sector of the model has an anomalous
$U(1)$ gauge symmetry, $U(1)_A$. This anomaly is cancelled by a Green--Shwarz
counterterm which induces a Fayet--Iliopoulos term into the D constraints
for $U(1)_A$ [\DSW]. The set of F and D constraints is given by the
following equations:
$$\eqalignno{&D_A=\sum_k Q^A_k \vert \chi_k \vert^2={-g^2e^{\phi_D}
\over 192\pi^2}Tr(Q_A) {1\over {2\alpha^{\prime}}} {\hskip .1cm},&(11a) \cr
&D^{\prime j}=\sum_k Q^{\prime j}_k \vert \chi_k \vert^2=0 \qquad j=1
\ldots 5  {\hskip .1cm},&(11b) \cr
&D^j=\sum_k Q^j_k \vert\chi_k\vert^2=0 \qquad j=C,L,7,8{\hskip .1cm},&(11c) \cr
&W={\partial W\over \partial \eta_i} =0 {\hskip .1cm}, &(11d) \cr}$$
where $\chi_k$ are the fields that get VEVs and $Q^j_k$ are their charges.
$W$ is the cubic superpotential and $\eta_i$ are the fields which do not get
VEVs. $\alpha^{\prime}$ is the string tension and $Tr(Q_A)=180$ in this model.
Eq. (11a) is the D constraint for the anomalous $U(1)_A$. We see that some
$SO(10)$ singlet scalars must get Planck scale VEVs in order to satisfy (11a)
and preserve SUSY around the Planck scale. Then, due to the other F and D
constraints most or all of the
other scalars also obtain VEVs. Since these scalars in general have nonzero
R charges all three R charges are spontaneously broken around the
Planck scale. For example, in the model of Ref. [\MOD] under consideration,
$\Phi_{45}$
must get a VEV in order to satisfy Eq. (11a). This and other VEVs of scalars
coming from the sector $b_1+b_2+\alpha+\beta$ break $Q_{\ell_1}$ and
$Q_{\ell_2}$ spontaneously around the Planck scale. SUSY F constraints in
the observable sector require that $\l \Phi_{12} \r=\l \bar \Phi_{12} \r=
\l \xi_3 \r=0$ [\NRT]. Thus, $Q_{\ell_3}$ can only be broken by hidden sector
VEVs. For example, $\l V_1 \r$ and $\l \bar V_2 \r$ which are needed to get
quark mixing break $Q_{\ell_3}$ spontaneously around the Planck scale [\CKM].

After the three $U(1)$ R symmetries are broken explicitly and spontaneously,
there still remains a discrete R symmetry in the model. This is a $Z_2$
subgroup
(of $U(1)^3$), i.e. an R parity. Charges of states under this R parity are
given by $exp(i \pi Q_R)$ where $Q_R=2\sum Q_{\ell_1}+Q_{\ell_2}+Q_{\ell_3}$.
The $Z_2$ charge
defined this way is invariant under picture changing since every picture
changing operation changes the sum by one and $Q_R$ by two. $Q_R$ is not broken
spontaneously either since all $SO(10)$ singlet scalars which get VEVs have
$Q_R=0,2$. We find that under this $Z_2$ all
superfields are even and $\theta$ is odd. As a result, all scalars are
even and all fermions are odd. This is equivalent to fermion
number modulo 2 and therefore does not give any new constraints on the model.
In particular this $Z_2$ symmetry which is not the usual R parity
(under which matter fermions are even) does not eliminate the dangerous terms
in Eq. (1).

In light of this result, how can the terms in Eq. (1) be suppressed in
standard--like superstring models?
In these models $B-L$ is gauged and given by $U(1)_{B-L}
=2 U(1)_C/3$. Conservation of $U(1)_{B-L}$ (or equivalently of $U(1)_C$)
eliminates the terms in Eq. (1) since they violate $B-L$ in addition to $B$
and $L$. Order $N>3$
terms which can induce these operators by scalar VEVs exist but they are all
proportional to $\l \tilde N_i \r$, the VEVs of the right--handed sneutrinos.
Explicitly the nonrenormalizable terms in the superpotential are [\NRT]
$$\eqalignno{&(u_3d_3+Q_3L_3)d_2N_2 \Phi_{45} \bar \Phi_2^-, &(12a) \cr
             &(u_3d_3+Q_3L_3)d_1N_1 \Phi_{45} \Phi_1^+, &(12b) \cr
             &u_3d_2d_2N_3 \Phi_{45} \bar \Phi_2^- +u_3d_1d_1N_3 \Phi_{45}
\Phi_1^+, &(12c) \cr
             &Q_3L_1d_3N_1 \Phi_{45} \Phi_3^+ +Q_3L_1d_1N_3 \Phi_{45} \Phi_3^+,
&(12d) \cr
             &Q_3L_2d_3N_2 \Phi_{45} \bar \Phi_3^- +Q_3L_2d_2N_3 \Phi_{45}
\bar \Phi_3^-, &(12e) }$$
for the (effective dimension four) $B$ violating operators. $L$ violating
operators have the same generic form and will not be written explicitly.

Now, one might think that, in these models, since $B-L$ is gauged, as long as
it is not spontaneously broken by $\l \tilde N_i \r$ the dangerous terms in
Eq. (1) are eliminated. The situation is more
complicated since there are hidden sector states with nonzero $B-L$ charge
which can get VEVs. In the notation of Ref. [\MOD], these are $H_i$ $i=14,
\ldots ,26$ with $Q_{B-L}=\pm 1/2$. SUSY F constraints in the hidden
sector require most of the VEVs of $H_i$ to vanish [\LQ]. Still, either
$H_{23}, H_{25}$ or $H_{24}, H_{26}$ may
get VEVs and break $U(1)_{B-L}$ spontaneously near the Planck scale. In
addition, the pair $H_{24},H_{26}$ has the correct $Q_{B-L}$ charge to render
the terms of Eq. (1) neutral under $B-L$. There is no danger of having $B$ (or
$L$) violating terms containing $H_{24}H_{26}$ however, due to conservation
of local $U(1)_{r_{j+3}}$. (Strictly speaking, this is true only when $H_{19}$
and $H_{20}$ get Planck scale masses and decouple. In these models, they get
masses of $10^{17}~GeV$ from the cubic superpotential [\MOD].) Therefore,
even though gauged $B-L$ can be broken by hidden sector VEVs, the terms in
Eq. (12a-e) are still
the only ones to induce those in Eq. (1). This is due to a conspiracy between
the $U(1)_{B-L}$ and $U(1)_{r_{j+3}}$ charges of the massless states in the
observable and hidden sectors. One can
think of this as an effective R parity of the model which can only be broken
by nonvanishing $\l \tilde N_i \r$. It is this effective R parity which
requires the
states $N_i$ to appear in all the $B$ (and $L$) violating terms in Eqs.
(12a-e). The magnitude of the coefficients $c_i$ in Eq. (1)
is controlled by $\l \tilde N_i \r$. Since there are no other constraints on
$\l \tilde N_i \r$ (from SUSY etc.), one can choose them to be zero or small
enough to satisfy the constraints from proton lifetime (and $L$ violating
processes). Reversing the argument, proton lifetime gives an upper bound
on $\l \tilde N_i \r$. The strongest constraints arise from the terms in
Eq. (12a,b) and give $\l \tilde N_i \r/M <10^{-11}$ or $\l \tilde N_i \r <
10^7~GeV$ for TeV scale squarks.

To summarize, we find that there are three continous $U(1)$ R symmetries
at the cubic level of the superpotential. These are broken explicitly by higher
order ($N>3$) terms in the superpotential and spontaneously by the scalar VEVs
which are necessary in order to preserve SUSY around the Planck scale. Then,
only an R parity ($Z_2$) which is is equivalent to fermion number modulo 2
survives.
On the other hand, standard--like superstring models possess an effective
R parity which arises from the particular charges of observable and hidden
sector under gauged $U(1)_{B-L}$ and $U(1)_{r_{j+3}}$.

The effective R parity which can be broken only by $\l \tilde N_i \r$ is not
enough to rule out large $B$ (or $L$) violation in standard--like superstring
models. There may be effective $B$ (or $L$) violating terms other than
the ones in Eq. (1). For example one must make sure that the effective $N=4$
terms in the superpotential such as
$$c_4 Q_i Q_j Q_k L_l+ c_5 u_i u_j d_k e_l + c_6 Q_i Q_j Q_k H_l \eqno(13)$$
which induce dimension five $B$ violating terms are sufficiently suppressed.
(Here $i,j,k,l$ are gereration indices.) In
addition, intermediate scale color triplets (such as the leptoquark
$D_{45}$ in standard--like superstring models [\LQ]) can induce effective
dimension five or six $B$ violating operators unless they have
weak enough couplings to matter fermions and/or large enough masses. A more
detailed study shows
that $B$ violation arising from the sources mentioned above can be suppressed
sufficiently in standard--like superstring models so that constraints from
proton lifetime are satisfied [\PRO] .

Arguments similar to the ones used in this letter can also be applied to the
other global symmetries of the cubic superpotential. For example, in the
model under consideration, the left--handed world--sheet currents
$y^3 y^6,y^1 \omega^5, \omega^2 \omega^4$ give three other global $U(1)$
symmetries of the cubic superpotential. These too are broken explicitly by
$N>3$ terms and spontaneously by the scalar VEVs and may lead to discrete
symmetries. In addition, string selection rules together with the specific
charges of massless states may result in discrete symmetries which are not
subgroups of continous global symmetries.

\bigskip
\centerline {\bf Acknowledgements}

This work is supported by the Department of Particle Physics and a Feinberg
Fellowship.

\refout
\vfill
\eject

\end
\bye